\documentclass[runningheads]{llncs}

\usepackage[mobile]{eccv}

\usepackage[table]{xcolor}
\usepackage{colortbl}
\usepackage{multirow}
\usepackage{array}
\usepackage{booktabs} %

\usepackage{eccvabbrv}

\usepackage{graphicx}
\usepackage{booktabs}

\usepackage[accsupp]{axessibility}  %
\usepackage{algorithm}
\usepackage{algpseudocode} %

\algrenewcommand\algorithmicrequire{\textbf{Input:}}
\algrenewcommand\algorithmicensure{\textbf{Output:}}
\algrenewcommand{\algorithmiccomment}[1]{\hfill{\footnotesize$\triangleright$~#1}}

\usepackage[protrusion=true,expansion=false]{microtype}
\usepackage[pagebackref,breaklinks,colorlinks,citecolor=eccvblue]{hyperref}

\usepackage{orcidlink}

\begin{document}

\title{DuoTeach: Dual Role Self-Teaching for Coarse-to-Fine Decision Coordination in Vision--Language Models}

\titlerunning{DuoTeach for Coarse-to-Fine Decision Coordination}

\author{
Wei Yang\inst{1}\textsuperscript{*} \and
Yiran Zhu\inst{1,2}\textsuperscript{*} \and
Zilin Li\inst{4} \and
Xunjia Zhang\inst{1} \and
Jun Xia\inst{2,3}\textsuperscript{$\dagger$} \and
Hongtao Wang\inst{1}\textsuperscript{$\dagger$}
}

\authorrunning{W. Yang et al.}

\institute{
Department of Computer, North China Electric Power University, Baoding, China
\and
AIMS Lab, HKUST (GZ), Guangzhou, China
\and
HKUST, Hong Kong SAR, China
\and
School of Information and Intelligent Science, Donghua University, Shanghai, China
}

\maketitle

\begingroup
\renewcommand\thefootnote{}
\footnotetext{\textsuperscript{*} Co-first authors.}
\footnotetext{\textsuperscript{$\dagger$} Co-corresponding authors: wanght@ncepu.edu.cn, junxia@hkust-gz.edu.cn.}
\endgroup
\setcounter{footnote}{0}

\begin{abstract}
Coarse-to-fine path decision-making requires predicting a valid taxonomy path whose earlier decisions constrain later ones. However, existing benchmarks score each level independently, obscuring cross-level validity and consistency. To align evaluation with this setting, we introduce a \emph{Joint Path Decision} (JPD) protocol that requires predicting the full path in one call, together with \emph{Depth-Weighted Prefix Accuracy} (DWPA), a metric family for measuring path reliability with tunable emphasis on deeper levels. Under JPD, strong vision--language models (VLMs) frequently produce invalid parent--child pairs and brittle full-path predictions, indicating that their failures stem not only from incomplete taxonomic knowledge but also from unstable cross-level decision coordination.
We address this with \emph{DuoTeach}, a dual role self-teaching distillation framework that requires no ground-truth labels and reuses the same pretrained VLM in two roles. Its \emph{Decision-Conditioned Rollout} (DCR) generates more coherent teacher traces by conditioning each level on prior decisions, and distills this coordinated behavior into the student without extra test-time rollouts.
Across multiple taxonomy-structured benchmarks and VLM base models, DuoTeach improves in-domain DWPA($\alpha{=}0.95$) by up to +30.24 points and boosts zero-shot performance on unseen taxonomies from 17.17\% to 43.66\%. Analyses attribute these gains to improved within-call multi-level decision coordination. Code will be released upon acceptance.
\end{abstract}

\section{Introduction}
\label{sec:intro}

Many applications require a model to return a fine-grained label together with a valid ancestry trail in a taxonomy (e.g., product categorization with breadcrumb paths~\cite{Lin_2019_ICTIR_ProductCat,Brinkmann_2021_DEB_Product}, medical coding with hierarchical code systems~\cite{Cao_2019_AMIA_ICD}, and safety taxonomies for moderation/routing~\cite{Li_2025_arXiv_HiGuard}). We refer to this setting as \emph{coarse-to-fine path decision-making}.
Given an input, the model makes one decision at each level, and the decisions must form a valid parent--child chain in the taxonomy. A fine-grained decision is only meaningful when its higher-level ancestors are correct, so evaluation should measure whether the model can produce a correct and coherent path, not just isolated per-level hits.

\begin{figure}[t]
  \centering
  \includegraphics[trim=50pt 185pt 50pt 145pt, clip, width=0.90\linewidth]{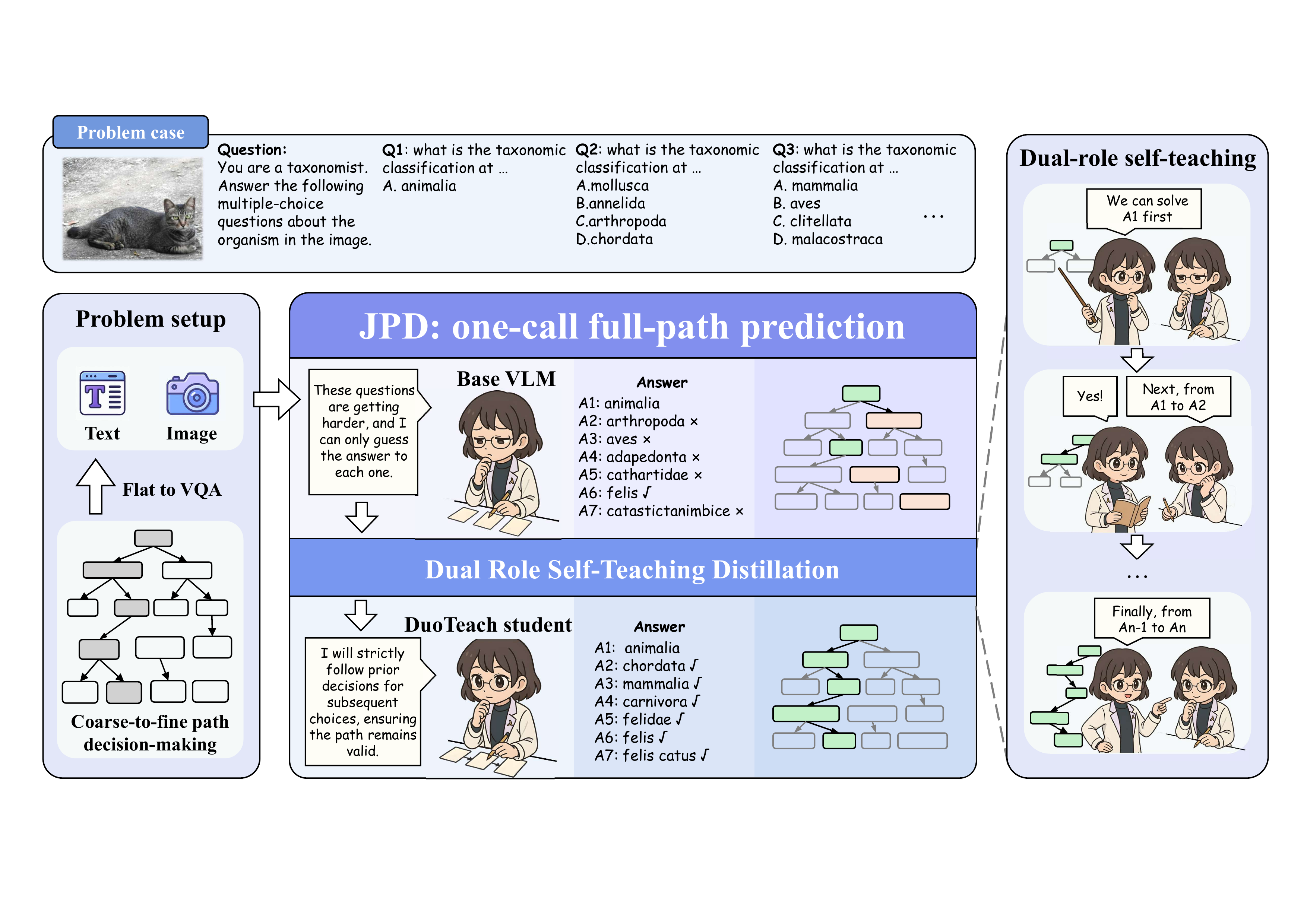}
  \caption{Overview of coarse-to-fine path decision-making under JPD and the DuoTeach framework.}
  \label{fig:intro}
\end{figure}

However, hierarchical benchmarks built from common taxonomic datasets~\cite{Deng_2009_CVPR_ImageNet,VanHorn_2021_CVPR_iNat2021,Wah_2011_TR_CUB200_2011,Bossard_2014_ECCV_Food101,Tan_2025_arXiv_HVU,Park_2025_ICLR_HCAST} are often evaluated in a level-wise manner, where each level is queried as an independent question and scored in isolation. Such a protocol does not enforce parent--child dependency: the model is not required to condition lower-level decisions on its own higher-level outputs within the same call. Consequently, this evaluation mainly measures per-level correctness under separate queries, rather than coarse-to-fine path decision-making, which requires coordinating a valid taxonomy path in one call. For example, a model may answer \emph{Animal} at L1 and \emph{Cat} at L3 correctly when asked separately, but output an inconsistent path such as \emph{Animal} $\rightarrow$ \emph{Bird} $\rightarrow$ \emph{Cat} when predicting jointly. Thus, level-wise scores can be optimistic about joint-path reliability.

To evaluate \emph{coarse-to-fine path decision-making}, we introduce a \emph{Joint Path Decision} (JPD) protocol and a new metric family, \emph{Depth-Weighted Prefix Accuracy} (DWPA). JPD requires predicting the full root-to-leaf path in one call, so lower-level decisions are made under the model's own higher-level outputs. DWPA measures path reliability via prefix correctness, with depth-increasing weights controlled by a single parameter $\alpha$ that sets the emphasis on the finest level. This yields two representative operating points: DWPA$_{0.95}$ for settings that demand strict leaf correctness, and DWPA$_{1/L}$ for measuring the depth of the longest correct prefix. Under JPD, strong VLMs still exhibit frequent inconsistent paths and large reliability drops; for example, on iNat-Animal with LLaVA-OV-7B, DWPA$_{0.95}$ is only 0.07\%. This shows that failures reflect not only incomplete taxonomic knowledge but also unstable cross-level decision coordination, favoring coordination improvements over taxonomy-specific supervision.

To reduce inconsistent paths under JPD, we propose \emph{DuoTeach}, a dual role self-teaching distillation framework that improves cross-level decision coordination without adding taxonomy-specific supervision. DuoTeach first uses the same frozen pretrained VLM to elicit teacher traces with a \emph{Decision-Conditioned Rollout} (DCR). DCR is a compute-intensive, multi-step level-by-level rollout that conditions each decision on previously predicted higher-level decisions, producing more coherent coarse-to-fine paths. We then distill this coordinated behavior into a one-call student by aligning teacher and student at the decision points, enabling coordinated cross-level decisions in a single call and reducing parent--child conflicts without extra test-time rollouts.

Experiments across taxonomy-structured benchmarks and VLM base models show that \emph{DuoTeach} improves performance under JPD consistently across all reported metrics, from path-level reliability (e.g., DWPA) to standard accuracy measures (e.g., leaf accuracy). It yields strong transfer to an unseen taxonomy (Food-101 held out during training), improving zero-shot DWPA$_{0.95}$ from 17.17\% to 43.66\%; we also see gains on external compositional VQA and mathematical reasoning benchmarks, suggesting potential transfer beyond a single label tree. Finally, diagnostic analyses of inconsistent paths and parent--child conflicts link DuoTeach's gains to more stable cross-level decision coordination under JPD.

In summary, our main contributions are as follows:
\begin{itemize}
    \item We introduce a JPD protocol for \emph{coarse-to-fine path decision-making}, which requires predicting the full taxonomy path in one call, together with DWPA, a metric family that measures path reliability via prefix correctness with tunable emphasis on deeper levels.

    \item Re-evaluating strong VLMs under JPD reveals frequent inconsistent paths and parent--child conflicts, showing that level-wise evaluation can be overly optimistic. Our analyses suggest that these failures are driven not only by per-level accuracy limits, but also by unstable cross-level decision coordination within a single call.

    \item We propose \emph{DuoTeach}, a dual role self-teaching distillation framework that uses the same frozen pretrained VLM to generate coherent teacher traces via a compute-intensive DCR, and distills this behavior into a one-call student by aligning at decision points. DuoTeach requires no external teacher model or extra test-time rollouts, and improves performance consistently across all reported metrics on multiple benchmarks and VLM backbones.
\end{itemize}

\section{Related Work}
\label{sec:related}

\paragraph{Taxonomy-Aware Methods for Coarse-to-Fine Paths.}
Most work that targets hierarchical predictions makes the taxonomy explicit in training or model design. Examples span constraint-based losses~\cite{Park_2025_ICLR_HCAST}, structure-preserving label spaces such as hyperbolic or graph embeddings~\cite{Xu_2023_NeurIPS_Hyperbolic,Yi_2022_ECCV_HGRNet,Noor_2025_KBS_TaxonomyRouting}, and refinement modules that couple coarse and fine outputs~\cite{Liu_2022_ECCV_WhereToFocus,Wang_2023_ACMMM_CAFL}. These choices can improve hierarchy compliance, but they usually tie the solution to a specific taxonomy and training signal. Our focus is different: we keep the base VLM unchanged and study \emph{coarse-to-fine path decision-making} under one-call prediction, evaluated with JPD and DWPA, where the key failure is inconsistent paths rather than isolated per-level errors.

\paragraph{Elicitation with Extra Test-Time Computation.}
Another line of work improves performance by adding computation at inference time. Multi-step prompting and tool- or agent-based pipelines expand the context with extra steps or external modules (e.g., retrieval, calculators)~\cite{Hu_2024_CVPR_VPD,Yin_2025_ICCV_ToolVQA}. Step-by-step supervision is also used to teach intermediate reasoning~\cite{Lu_2022_NeurIPS_ScienceQA}, but it requires process labels. DuoTeach follows the elicitation idea without adding inference overhead: it uses DCR to generate better traces only during training, then runs in one call at test time.

\paragraph{Distillation into One-Call Inference.}
Distillation is a standard way to compress costly behavior into a single forward pass. Classical self-distillation aligns outputs or features within a model~\cite{Furlanello_2018_ICML_BAN,Zhang_2019_ICCV_BYOT,Zhang_2020_NeurIPS_SelfDistill}, but it is not tailored to cross-level dependencies along a taxonomy path. More recent VLM pipelines distill tool-augmented or multi-step traces into a single model~\cite{Hu_2024_CVPR_VPD,Qin_2025_SciRep_EKDA,Shi_2025_CVPR_AoTD}, often using external teachers, tools, or explicit language chains. DuoTeach instead uses a single frozen VLM in a dual-role setup: DCR elicits coherent traces, and a student matches the teacher at decision points,. This directly targets cross-level decision coordination under JPD, without external teacher models or extra test-time rollouts.

\section{Approach}
\label{sec:approach}
We study \emph{coarse-to-fine path decision-making}, where a model must output a valid taxonomy path in one call. To evaluate this setting, we introduce a \emph{Joint Path Decision} (JPD) protocol and a new metric family, \emph{Depth-Weighted Prefix Accuracy} (DWPA). Under JPD, strong VLMs still exhibit frequent inconsistent paths, motivating our method \emph{DuoTeach}, a dual role self-teaching distillation framework that improves cross-level decision coordination

\subsection{Task and Protocol (JPD)}
\label{sec:task_jpd}
We study \emph{coarse-to-fine path decision-making} over a fixed taxonomy tree $\mathcal{T}$ of depth $L$.
For each image $x$, the target is a category sequence $\mathbf{y}_{1:L}=(y_1,\dots,y_L)$, where $y_\ell$ denotes the level-$\ell$ category.
Each level $\ell$ is associated with a multiple-choice query $q_\ell$ with an option set $\mathcal{A}_\ell$.
We denote by $\mathrm{Tok}(y_\ell)$ the surface option token used in prompts (e.g., ``A'').
Given a prediction $\hat{\mathbf{y}}_{1:L}=(\hat{y}_1,\dots,\hat{y}_L)$, it is \emph{valid} if the labels form a root-to-leaf ancestry chain in $\mathcal{T}$.

Under JPD, the model receives the image together with all level-specific queries $(q_1,\dots,q_L)$ in a single call, and outputs a label sequence $\hat{\mathbf{y}}_{1:L}$ by selecting one option per level.
The selected options correspond to predicted labels $\hat{\mathbf{y}}_{1:L}$, which must form a valid coarse-to-fine path in $\mathcal{T}$.
This protocol contrasts with level-wise evaluation that queries and scores each level in isolation, and it directly reflects the requirement of producing a coherent full path in one call.

\begin{figure}[t]
  \centering
  \includegraphics[trim=0 230pt 0 230pt, clip, width=0.98\linewidth]{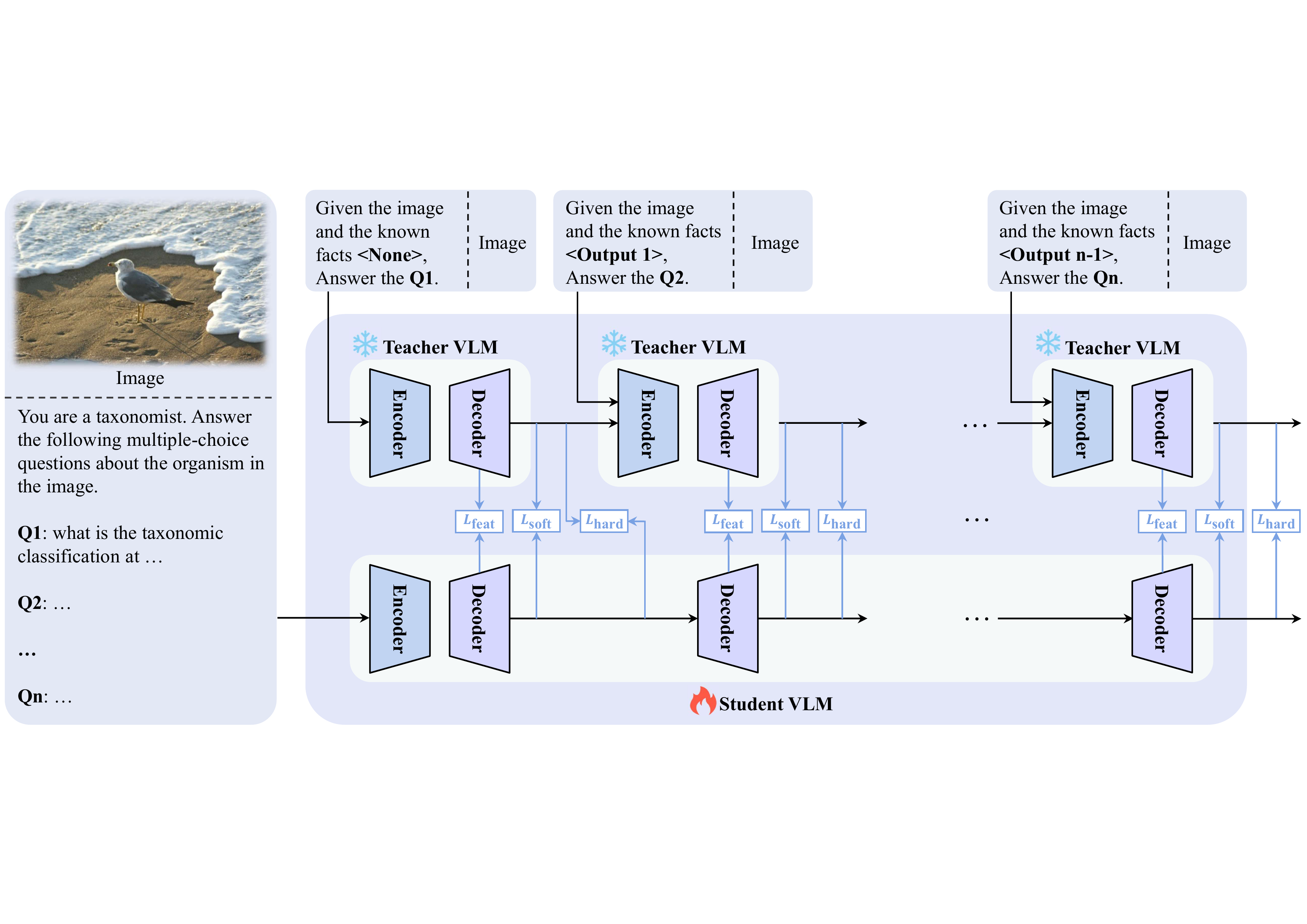}
  \caption{DuoTeach overview: the same pretrained VLM plays two roles. A frozen teacher uses a compute-intensive DCR to generate coherent coarse-to-fine decision traces, which are distilled into a one-call student for JPD inference.}
  \label{fig:method}
\end{figure}

\subsection{Metrics}
\label{sec:metrics}
We measure path reliability with \emph{Depth-Weighted Prefix Accuracy} (DWPA). For a sample with hierarchy depth $L$, let the ground-truth path be $\mathbf{y}=(y_1,\dots,y_L)$ and the model prediction be $\hat{\mathbf{y}}=(\hat{y}_1,\dots,\hat{y}_L)$. We first define the prefix-correctness indicator
\begin{equation}
c_\ell \;=\; \mathbb{I}\big[ \forall j\le \ell,\ \hat{y}_j = y_j \big],\qquad \ell=1,\dots,L,
\end{equation}
which credits depth $\ell$ only when all earlier coarse-to-fine decisions up to $\ell$ are also correct. DWPA is then a depth-weighted sum of these indicators:
\begin{equation}
\mathrm{DWPA}_\alpha(\hat{\mathbf{y}},\mathbf{y})
\;=\;
\sum_{\ell=1}^{L} w_\ell^{(\alpha,L)}\, c_\ell,
\qquad
\sum_{\ell=1}^{L} w_\ell^{(\alpha,L)}=1,\ \ w_\ell^{(\alpha,L)}\ge 0.
\label{eq:dwpa_def}
\end{equation}
To reflect the increasing value of reaching deeper correct decisions, we use weights that grow monotonically with depth via a geometric form:
\begin{equation}
w_\ell^{(\alpha,L)}
\;=\;
\frac{r_{\alpha,L}^{\,\ell-1}}{\sum_{t=1}^{L} r_{\alpha,L}^{\,t-1}},
\qquad r_{\alpha,L}\ge 1,
\label{eq:dwpa_weights}
\end{equation}
where the growth rate $r_{\alpha,L}$ is chosen such that the finest level receives a prescribed mass,
\begin{equation}
w_{L}^{(\alpha,L)}=\alpha.
\label{eq:dwpa_alpha}
\end{equation}
Equations~\eqref{eq:dwpa_weights}--\eqref{eq:dwpa_alpha} uniquely determine $r_{\alpha,L}$ for any $(\alpha,L)$. We use two representative operating points. DWPA$_{0.95}$ sets $\alpha=0.95$, prioritizing correctness at the finest level while still assigning progressively larger credit to deeper correct prefixes. DWPA$_{1/L}$ sets $\alpha=1/L$, which yields $r_{\alpha,L}=1$ and uniform weights $w_\ell^{(1/L,L)}=1/L$; in this case,
\begin{equation}
\mathrm{DWPA}_{1/L} \;=\; \frac{1}{L}\sum_{\ell=1}^L c_\ell \;=\; \frac{k}{L},
\end{equation}
where $k$ is the number of consecutive correct decisions from the root (i.e., the correct-prefix depth). We additionally report \emph{LeafAcc}, the accuracy at the finest level, and \emph{TOR} as a complementary hierarchical metric (see Appendix B for details).

\begin{algorithm}[t]
\caption{DuoTeach training pipeline}
\label{alg:duoteach}
\small
\begin{algorithmic}[1]
\Require Image $x$; level queries and options $\{q_\ell,\mathcal{A}_\ell\}_{\ell=1}^L$; frozen teacher $t_\phi$; student $s_\theta$
\Ensure Updated student parameters $\theta$

\State \textbf{DCR teacher rollout:} $A \gets \emptyset$ \Comment{accumulated prior decision tokens}
\For{$\ell \gets 1$ to $L$}
  \State $c_\ell \gets \mathrm{Input}(x, q_\ell, A)$ \Comment{append prior decision tokens to the prompt}
  \State $(\tilde{y}_\ell,\;p_\ell^{(T)},\;h_\ell^{(T)}) \gets t_\phi(c_\ell)$ \Comment{hard option / probs / decision-point state}
  \State $A \gets A \cup \{\mathrm{Tok}(\tilde{y}_\ell)\}$
\EndFor

\State \textbf{JPD student inference:} $c \gets \mathrm{Input}(x, \{q_\ell\}_{\ell=1}^L)$
\State $\{(p_\ell^{(S)}, h_\ell^{(S)})\}_{\ell=1}^L \gets s_\theta(c)$ \Comment{read per-level states at anchored decision points}

\State $\mathcal{L} \gets \sum_{\ell=1}^L \Big[
\lambda_1\,\mathrm{CE}(p_\ell^{(S)},\tilde{y}_\ell)
+\lambda_2\,\mathrm{KL}(p_\ell^{(T)}\|p_\ell^{(S)})
+\lambda_3\,\|W h_\ell^{(S)}-h_\ell^{(T)}\|_2^2 \Big]$
\State Update $\theta$; keep $\phi$ frozen
\end{algorithmic}
\end{algorithm}

\subsection{DuoTeach: Dual Role Self-Teaching Distillation}
\label{sec:duoteach}
We propose \emph{DuoTeach}, a dual role self-teaching distillation framework for improving cross-level decision coordination under JPD. DuoTeach reuses the same pretrained VLM in two roles: a frozen teacher $t_\phi$ that generates coherent coarse-to-fine decision traces via a compute-intensive \emph{Decision-Conditioned Rollout} (DCR), and a trainable one-call student $s_\theta$ that distills these traces to make coordinated path predictions under JPD without extra test-time rollouts (Fig.~\ref{fig:method}).

\paragraph{DCR teacher rollout.}
We use the frozen teacher $t_\phi$ to elicit level-wise decision traces during training. Under DCR, the teacher answers levels sequentially and conditions each level-$\ell$ decision on its previously generated decisions by appending them to the prompt as fixed context. Given an image $x$ and queries $(q_1,\dots,q_L)$, the teacher input at level $\ell$ is
\begin{equation}
c_\ell=\mathrm{Input}\big(x, q_\ell, \mathrm{Tok}(\tilde{y}_{1:\ell-1})\big),
\end{equation}
where $\mathrm{Tok}(\tilde{y}_{1:\ell-1})$ are the teacher's textual decisions from levels $1$ to $\ell{-}1$ (empty for $\ell{=}1$). A forward pass on $c_\ell$ yields an option distribution $p_\ell^{(T)}$ over $\mathcal{A}_\ell$, and we take the hard decision
\begin{equation}
\tilde{y}_\ell=\arg\max_{\alpha\in\mathcal{A}_\ell} p_\ell^{(T)}(\alpha),
\end{equation}
then append the corresponding option token $\mathrm{Tok}(\tilde{y}_\ell)$ to form the context for the next level.
To transfer decision states, we extract the final-layer hidden state at the anchored decision point, i.e., the generated token that expresses the selected option (e.g., the letter token ``A''). Let $H_\ell^{(T)}=(h_1,\dots,h_{T_\ell})$ be the decoder hidden states when generating the level-$\ell$ answer, and let $\mathrm{anchor}_\ell$ denote the position of the option token $\mathrm{Tok}(\tilde{y}_\ell)$. We define
\begin{equation}
h_\ell^{(T)} = h_{\mathrm{anchor}_\ell}.
\end{equation}
Running $\ell=1\to L$ produces teacher traces $\{(\tilde y_\ell, p^{(T)}_\ell, h^{(T)}_\ell)\}_{\ell=1}^L$.

\paragraph{One-call student under JPD.}
The student $s_\theta$ is initialized from the same pretrained VLM and trained to operate under JPD (Sec.~\ref{sec:task_jpd}): it receives the image and all level-specific queries in a single call and must maintain cross-level coordination without injected prior decisions. Decoding is autoregressive; letting $\hat{y}_\ell$ denote the option predicted at level $\ell$,
\begin{equation}
\hat{y}_\ell \sim P_\theta(\cdot \mid x, q_{1:L}, \hat{y}_{1:\ell-1}), \quad \ell = 1,\dots,L,
\end{equation}
where $P_\theta$ is the conditional distribution defined by $s_\theta$. At each level $\ell$, the student yields an option distribution $p_\ell^{(S)}$ over $\mathcal{A}_\ell$ and a hard decision $\hat{y}_\ell$. Using the same anchored decision point definition as the teacher, we read the student decision state $h_\ell^{(S)}$ at $\mathrm{anchor}_\ell$, producing $\{(p^{(S)}_\ell, h^{(S)}_\ell)\}_{\ell=1}^L$ aligned with the teacher traces.

\paragraph{Distillation at decision points.}
We distill DCR teacher traces into the one-call student by aligning teacher and student at the anchored decision points for each level. For level $\ell$, we use three complementary losses:
\begin{align}
\mathcal{L}_{\mathrm{hard}} &= \sum_{\ell=1}^{L} \mathrm{CE}\big(p^{(S)}_\ell, \tilde{y}_\ell\big), \\
\mathcal{L}_{\mathrm{soft}} &= \sum_{\ell=1}^{L} \mathrm{KL}\big(p^{(T)}_\ell \,\|\, p^{(S)}_\ell\big), \\
\mathcal{L}_{\mathrm{feat}} &= \sum_{\ell=1}^{L} \big\| W h_\ell^{(S)} - h_\ell^{(T)} \big\|_2^2,
\end{align}
where $W$ projects student features to the teacher feature space. The overall objective is
\begin{equation}
\mathcal{L}
= \lambda_{1}\mathcal{L}_{\mathrm{hard}}
+ \lambda_{2}\mathcal{L}_{\mathrm{soft}}
+ \lambda_{3}\mathcal{L}_{\mathrm{feat}},
\end{equation}
with loss weights $\lambda_1,\lambda_2,\lambda_3$. By focusing supervision on the anchored decision points, the student learns to reproduce coordinated cross-level decisions in a single call.

\paragraph{Training pipeline.}
Algorithm~\ref{alg:duoteach} summarizes DuoTeach. We first run DCR with the frozen teacher to obtain per-level decisions, option distributions, and decision-point states. We then run the student once under JPD and optimize the distillation objective above.

\section{Experiments}
\label{sec:experiments}
We evaluate DuoTeach under JPD with DWPA as the primary metric. We study (1) distillation effectiveness from DCR to a one-call student, (2) key failure factors and ablations, and (3) transfer and practicality across model sizes.

\begingroup
\arrayrulecolor{black}
\renewcommand{\arraystretch}{1.1}
\setlength{\tabcolsep}{5pt}
\setlength{\arrayrulewidth}{0.4pt}

\definecolor{HeadRow}{HTML}{E2E1FF} %
\definecolor{RowA}{HTML}{FFFFFF}
\definecolor{RowB}{HTML}{F6FAFC}

\begin{table}[t]
  \caption{Dataset statistics. \# Levels: depth of hierarchy; \# Items: total items in the benchmark paper.}
  \label{tab:dataset_stats}
  \centering
  \small

  \begin{tabular}{%
    >{\hspace{0pt}}l<{\hspace{2pt}} !{\color{black}\vrule width 0.3pt} %
    >{\hspace{2pt}}l<{\hspace{2pt}} !{\color{black}\vrule width 0.3pt} %
    >{\hspace{2pt}}c<{\hspace{2pt}} r<{\hspace{2pt}}@{}}              %

    \noalign{\global\arrayrulewidth=0.8pt}\hline
    \noalign{\global\arrayrulewidth=0.4pt}
    \noalign{\vskip 1pt}

    \multicolumn{1}{>{\columncolor{HeadRow}}l<{\hspace{2pt}}!{\color{black}\vrule width 0.3pt}}{\rule{0pt}{3ex}\textbf{Type}} &
    \multicolumn{1}{>{\columncolor{HeadRow}\hspace{2pt}}l<{\hspace{2pt}}!{\color{black}\vrule width 0.3pt}}{\textbf{Dataset}} &
    \multicolumn{1}{>{\columncolor{HeadRow}\hspace{2pt}}c<{\hspace{2pt}}}{\textbf{\# Levels}} &
    \multicolumn{1}{>{\columncolor{HeadRow}}r}{\textbf{\# Items}} \\

    \noalign{\vskip 1pt}\hline\noalign{\vskip 1pt}
    \rowcolors{2}{RowA}{RowB}

    \multirow{6}{*}{\raggedright Internal} & CUB-200-2011~\cite{Wah_2011_TR_CUB200_2011}        & 4  & 5,794  \\
                                            & iNaturalist-Plant~\cite{VanHorn_2021_CVPR_iNat2021}  & 7  & 42.71K \\
                                            & iNaturalist-Animal~\cite{VanHorn_2021_CVPR_iNat2021} & 7  & 53.88K \\
                                            & ImageNet-Animal~\cite{Deng_2009_CVPR_ImageNet}       & 12 & 19.85K \\
                                            & ImageNet-Artifact~\cite{Deng_2009_CVPR_ImageNet}     & 8  & 24.55K \\
                                            & Food-101~\cite{Bossard_2014_ECCV_Food101}            & 5  & 21.00K \\
    \noalign{\vskip 1pt}\hline\noalign{\vskip 1pt}

    \multirow{3}{*}{\raggedright External}  & GQA (test-dev)~\cite{Hudson_2019_CVPR_GQA}  & -- & $\sim$113K \\
                                            & MathVista~\cite{Lu_2024_ICLR_MathVista}     & -- & 6,141 \\
                                            & MMBench~\cite{Liu_2024_ECCV_MMBench}        & -- & 2,974 \\
    \noalign{\vskip 1pt}
    \noalign{\global\arrayrulewidth=0.8pt}\hline
    \noalign{\global\arrayrulewidth=0.4pt}
  \end{tabular}
\end{table}

\rowcolors{0}{}{}
\endgroup

\subsection{Experimental Setup}
\label{sec:expsetup}

\textit{Datasets.}
We evaluate on taxonomy-structured datasets including iNaturalist (iNat-Animal, iNat-Plant)~\cite{VanHorn_2021_CVPR_iNat2021}, ImageNet (Animal, Artifact)~\cite{Deng_2009_CVPR_ImageNet}, CUB-200-2011~\cite{Wah_2011_TR_CUB200_2011}, and Food-101~\cite{Bossard_2014_ECCV_Food101}. 
For each dataset, we use its taxonomy to form $L$ level-wise multiple-choice queries and follow standard hierarchical benchmark constructions for level definitions and query formatting~\cite{Tan_2025_arXiv_HVU}. 
Unless stated otherwise, we split each internal dataset into train/val/test with a 6:2:2 ratio. 
We hold out Food-101 from training and evaluate it as an in-domain unseen taxonomy.
For external generalization, we report results on GQA and MathVista~\cite{Lu_2024_ICLR_MathVista}; for stability, we use MMBench~\cite{Liu_2024_ECCV_MMBench}. 
Dataset statistics are in Table~\ref{tab:dataset_stats}.

\textit{Evaluation Setting and Metrics.}
We evaluate taxonomy-structured datasets under JPD and report DWPA$_{0.95}$, DWPA$_{1/L}$, LeafAcc, and TOR. On non-taxonomy benchmarks, we report Acc. 

\textit{Implementation Details.}
We implement DuoTeach in PyTorch and run all experiments on NVIDIA L20 and H20 GPUs. We evaluate four pretrained VLM backbones: LLaVA-OV-7B~\cite{Li_2025_TMLR_LLaVAOneVision}, InternVL2.5-8B~\cite{Chen_2024_arXiv_ExpandingBoundaries}, InternVL3-8B~\cite{Zhu_2025_arXiv_InternVL3}, and Qwen2.5-VL-7B~\cite{Bai_2025_arXiv_Qwen2_5_VL}; \textit{Base} denotes the corresponding official checkpoint without adaptation. For all backbones, we freeze the vision encoder and fine-tune the language component with LoRA (rank $r{=}64$, $\alpha{=}16$, dropout $0.05$) applied to the Q/K/V/O projection layers. Images are resized to 448 pixels on the long side, followed by center-crop or padding. We train with AdamW (lr $2\times10^{-5}$, weight decay $0.01$, $\beta{=}(0.9,0.999)$), cosine decay with 3\% warmup, and FP16. Unless stated otherwise, we use $\lambda_{1}{=}2.0$, $\lambda_{2}{=}1.0$, $\lambda_{3}{=}0.5$, gradient clipping at $1.0$, and train up to 3 epochs with gradient accumulation (effective batch size 1). Additional details are in Appendix D.

\begingroup
\arrayrulecolor{black}
\setlength{\tabcolsep}{1.2pt}
\renewcommand{\arraystretch}{1.08}
\setlength{\arrayrulewidth}{0.3pt}

\definecolor{RowA}{HTML}{FFFFFF}
\definecolor{RowB}{HTML}{E7E9FF}
\definecolor{HeadText}{HTML}{000000}

\newcommand{\gain}[1]{{\fontsize{6pt}{6.6pt}\selectfont\textbf{(\,#1\,)}}}
\newcommand{\score}[2]{%
  \begin{tabular}[c]{@{}c@{}}#1\\[-0.5ex]\gain{#2}\end{tabular}%
}

\newcommand{\HeadY}{\color{black}\vrule width 0.3pt}
\renewcommand{\multirowsetup}{\raggedright}

\newcommand{\IconHLLAVA}{10pt}
\newcommand{\IconHInterVL}{9.5pt}
\newcommand{\IconHQwen}{12pt}
\definecolor{LLaVAColor}{HTML}{EA5A47}
\definecolor{InterVLColor}{HTML}{4D7DDE}
\definecolor{QwenColor}{HTML}{8A6BEA}

\newcommand{\ModelStack}[4]{%
  \makebox[\linewidth][c]{%
    \begin{tabular}[t]{@{}c@{}}
      \textcolor{#3}{\textbf{#4}}\\[-1pt]
      \colorbox{white}{\raisebox{-0.15\height}{\includegraphics[height=#1]{icon/#2}}}%
    \end{tabular}%
  }%
}

\begin{table*}[t]
  \caption{Main performance on the ImageNet-Animal test set. All values are percentages (\%). Parentheses show absolute gains over the Base model.}
  \label{tab:main_internal}
  \centering
  \scriptsize

  \begin{tabular}{@{}%
    >{\raggedright\arraybackslash}p{2.6cm} !{\HeadY}
    >{\centering\arraybackslash}p{1.9cm} !{\HeadY}
    cccc@{}}
    \noalign{\global\arrayrulewidth=0.8pt}\hline
    \noalign{\global\arrayrulewidth=0.3pt}

    \noalign{\vskip 2pt}
    \multicolumn{1}{>{\columncolor{RowB}}c!{\HeadY}}{\rule{0pt}{3.0ex}\textcolor{HeadText}{\textbf{Model}}} &
    \multicolumn{1}{>{\columncolor{RowB}\centering\arraybackslash}p{1.9cm}!{\HeadY}}{\textcolor{HeadText}{\textbf{Compare}}} &
    \multicolumn{1}{>{\columncolor{RowB}}c}{\textcolor{HeadText}{\textbf{DWPA$_{0.95}$}}} &
    \multicolumn{1}{>{\columncolor{RowB}}c}{\textcolor{HeadText}{\textbf{LeafAcc}}} &
    \multicolumn{1}{>{\columncolor{RowB}}c}{\textcolor{HeadText}{\textbf{TOR}}} &
    \multicolumn{1}{>{\columncolor{RowB}}c}{\textcolor{HeadText}{\textbf{DWPA$_{1/L}$}}\rule[-0.9ex]{0pt}{0pt}} \\
    \noalign{\vskip 2pt}\hline\noalign{\vskip 2pt}
    
    \multirow{3}{*}{\cellcolor{white}\ModelStack{\IconHLLAVA}{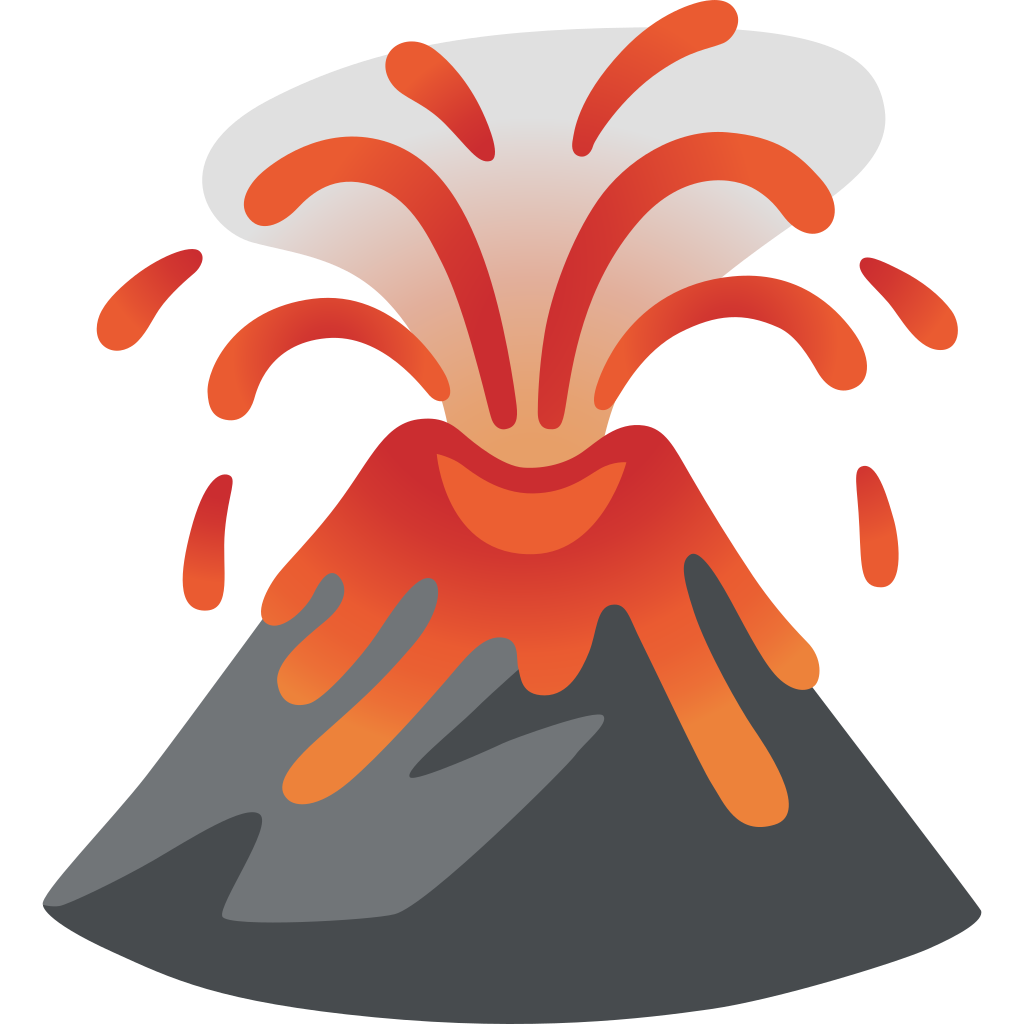}{LLaVAColor}{LLaVA-OV-7B}}
      & \textit{Base}    & 0.69 & 26.85 & 17.10 & 25.16 \\
      & \textit{Teacher} &
        \score{34.11}{+33.42} & \score{64.40}{+37.55} & \score{74.76}{+57.66} & \score{73.55}{+48.39} \\
      & \textit{DuoTeach} &
        \score{30.93}{+30.24} & \score{62.30}{+35.45} & \score{74.33}{+57.23} & \score{73.65}{+48.49} \\
    \noalign{\vskip 2pt}\hline\noalign{\vskip 2pt}

    \multirow{3}{*}{\ModelStack{\IconHInterVL}{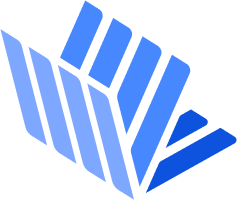}{InterVLColor}{InternVL2.5-8B}}
      & \textit{Base}    & 10.46 & 43.75 & 39.71 & 40.96 \\
      & \textit{Teacher} &
        \score{36.56}{+26.10} & \score{64.50}{+20.75} & \score{77.13}{+37.42} & \score{78.16}{+37.20} \\
      & \textit{DuoTeach} &
        \score{38.09}{+27.63} & \score{65.20}{+21.45} & \score{78.53}{+38.82} & \score{79.42}{+38.46} \\
    \noalign{\vskip 2pt}\hline\noalign{\vskip 2pt}

    \multirow{3}{*}{\ModelStack{\IconHInterVL}{InterVL.png}{InterVLColor}{InternVL3-8B}}
      & \textit{Base}    & 6.68 & 32.60 & 24.60 & 32.61 \\
      & \textit{Teacher} &
        \score{37.08}{+30.40} & \score{68.60}{+36.00} & \score{76.72}{+52.12} & \score{76.84}{+44.23} \\
      & \textit{DuoTeach} &
        \score{24.49}{+17.81} & \score{54.25}{+21.65} & \score{59.21}{+34.61} & \score{62.43}{+29.82} \\
    \noalign{\vskip 2pt}\hline\noalign{\vskip 2pt}

    \multirow{3}{*}{\ModelStack{\IconHQwen}{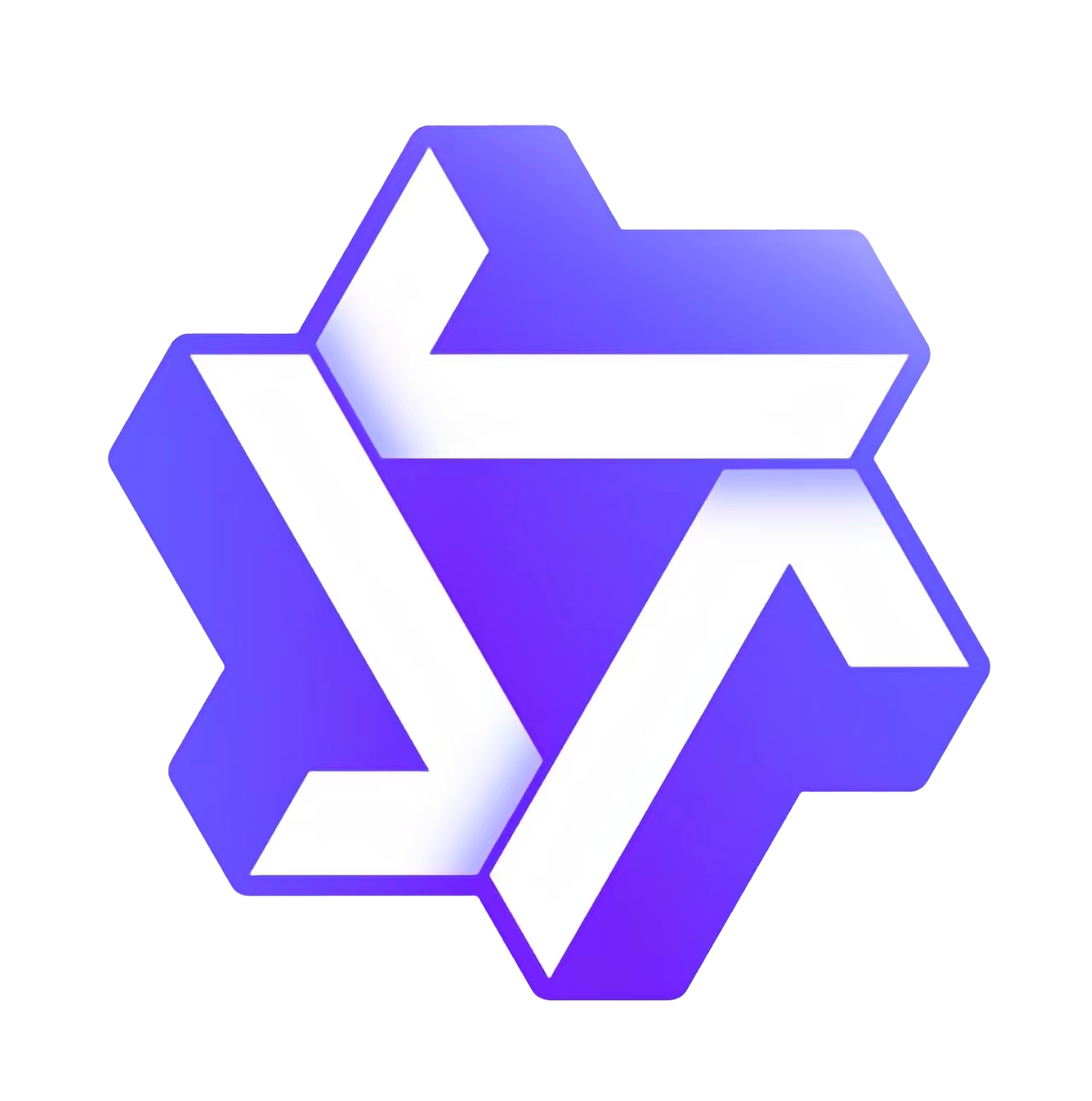}{QwenColor}{Qwen2.5-VL-7B}}
      & \textit{Base}    & 34.06 & 69.85 & 67.43 & 66.06 \\
      & \textit{Teacher} &
        \score{49.18}{+15.12} & \score{77.65}{+7.80}  & \score{82.57}{+15.14} & \score{82.38}{+16.32} \\
      & \textit{DuoTeach} &
        \score{50.31}{+16.25} & \score{78.70}{+8.85}  & \score{83.38}{+15.95} & \score{83.30}{+17.24} \\
    \noalign{\vskip 2pt}
    \noalign{\global\arrayrulewidth=0.8pt}\hline
    \noalign{\global\arrayrulewidth=0.3pt}
  \end{tabular}
\end{table*}

\rowcolors{0}{}{}
\endgroup

\subsection
{Main Results}
\label{main_result}
We evaluate four VLM backbones on ImageNet-Animal. Table~\ref{tab:main_internal} reports JPD results for the \textit{Base} and DuoTeach \textit{student}, plus a \textit{DCR teacher} upper bound.

Under JPD, the Base models largely fail to produce reliable paths: for example, LLaVA-OV-7B attains DWPA$_{0.95}$ of 0.69 and InternVL3-8B attains 6.68, despite non-trivial LeafAcc. Running the same frozen checkpoint as a multi-step DCR teacher substantially repairs this, yielding large absolute gains in DWPA$_{0.95}$ and transition-level consistency (e.g., LLaVA’s DWPA$_{0.95}$ rises to 34.11 and TOR to 74.76). Crucially, the DuoTeach student recovers most of this improvement while remaining one-call: for LLaVA, it reaches DWPA$_{0.95}$ 30.93 and TOR 74.33, with similar patterns for InternVL2.5-8B, InternVL3-8B, and Qwen2.5-VL-7B.
These improvements are not merely leaf-level; they coincide with systematic increases in TOR and DWPA$_{1/L}$, indicating more stable cross-level decision coordination. Overall, the Base--Teacher--Student comparison shows that DuoTeach internalizes compute-intensive DCR rollouts into efficient JPD inference. Further results are reported in Appendix F.1.

\begingroup
\setlength{\tabcolsep}{4pt}
\renewcommand{\arraystretch}{1.15}

\newcommand{\celltwo}[2]{%
  \begin{tabular}[c]{@{}c@{}}#1\\[-0.35ex]{\scriptsize(\,#2\,)}\end{tabular}%
}

\begin{table*}[t]
  \caption{Diagnostic comparison under three invocation modes (iNat-Animal, LLaVA-OV-7B).
  All metrics are percentages (\%). Parentheses indicate absolute change vs.\ JPD.
  More diagnostic results on more models and datasets are provided in Appendix E.}
  \label{tab:protocols}
  \centering
  \small

  \begin{tabular}{lcccc}
    \toprule
    Protocol &
    DWPA$_{0.95}$ &
    LeafAcc &
    TOR &
    DWPA$_{1/L}$ \\
    \midrule

    JPD
      & 0.07
      & 25.35
      & 11.92
      & 20.59 \\

    Independent Levels
      & \celltwo{4.99}{+4.92}
      & \celltwo{27.45}{+2.10}
      & \celltwo{51.92}{+40.00}
      & \celltwo{55.43}{+34.84} \\

    DCR
      & \celltwo{9.16}{+9.09}
      & \celltwo{32.45}{+7.10}
      & \celltwo{55.66}{+43.74}
      & \celltwo{57.71}{+37.12} \\

    \bottomrule
  \end{tabular}
\end{table*}

\endgroup

\subsection{Diagnosis: Why Paths Break under JPD}
\label{sec:diagnostic}
Main results show that several Base VLMs achieve non-trivial LeafAcc yet exhibit very low DWPA under JPD, suggesting that failures are dominated by cross-level coordination rather than leaf recognition alone. To isolate what drives this breakdown, we run a controlled protocol analysis: we fix the checkpoint, images, and query format, and vary only the invocation mode.

\textit{(i) Protocol ladder.}
We compare three inference modes.
(1) JPD: one call with all level-wise queries, producing one option per level.
(2) Independent Levels: each level is queried in a separate call and the answers are concatenated into a path.
(3) DCR: a stepwise rollout that queries level by level and appends the previous decision to the next-level prompt.

\begin{figure}[t]
  \centering
  \includegraphics[trim=0pt 10pt 0pt 10pt, clip, width=0.8\linewidth]{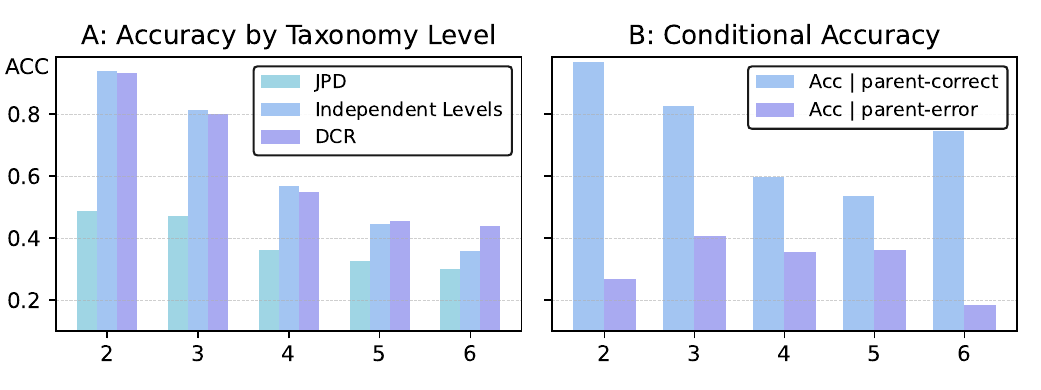}
  \caption{Depth-wise and conditional analysis across invocation modes.
  (A) Per-level accuracy vs.\ depth under JPD, Independent Levels, and DCR.
  (B) Parent--child coupling quantified by the conditional gap $\Delta_\ell$.}
  \label{fig:teaser}
\end{figure}

\begin{figure}[t]
  \centering
  \includegraphics[width=0.95\linewidth]{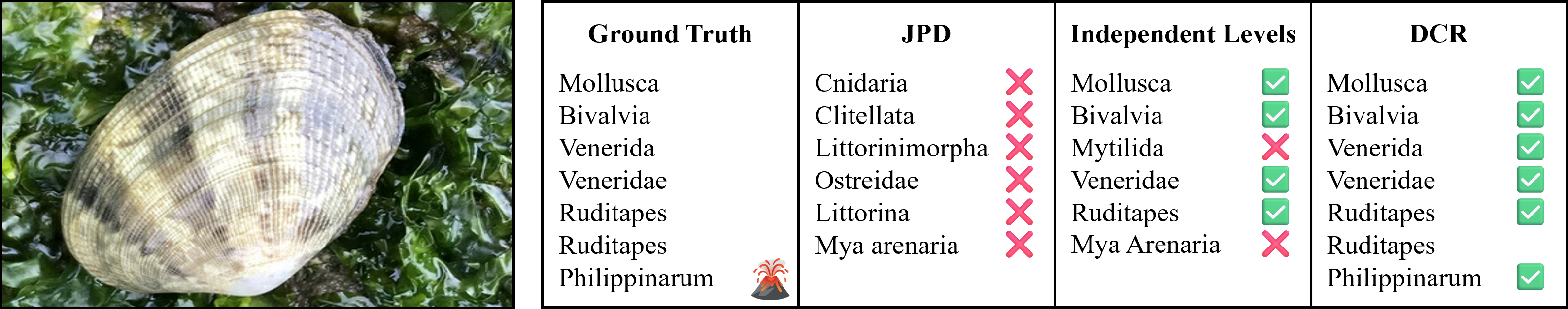}
  \caption{A qualitative example of path predictions under three inference modes. }
  \label{fig:case}
\end{figure}

\textit{(ii) Protocol comparison.}
Table~\ref{tab:protocols} shows a representative case (iNat-Animal, LLaVA-OV-7B). Under JPD, DWPA$_{0.95}$ nearly collapses (0.07), while LeafAcc is much higher (25.35), indicating that the model can often identify the leaf but fails to maintain a valid, consistent prefix. Independent Levels yields a clear improvement in path reliability and transition consistency (e.g., TOR rises from 11.92 to 51.92), showing that removing the single-call joint decoding burden recovers part of the performance. However, DWPA$_{0.95}$ remains low (4.99), suggesting that path coherence is still brittle.

DCR further improves DWPA$_{0.95}$ to 9.16 and increases TOR and DWPA$_{1/L}$. Since the only difference from Independent Levels is the explicit injection of prior decisions, the extra gain can be attributed to decision conditioning: propagating earlier decisions helps maintain coherent parent--child transitions beyond decomposition alone.

\textit{(iii) Where the gains occur.}
Figure~\ref{fig:teaser} decomposes these effects by depth. DCR provides larger benefits at deeper levels, where errors are more sensitive to upstream decisions. We quantify this sensitivity by comparing level-$\ell$ accuracy conditioned on whether its parent is correct:
\begin{equation}
\Delta_{\ell} = \mathrm{Acc}_{\ell}\mid \text{parent-correct} \;-\; \mathrm{Acc}_{\ell}\mid \text{parent-error}.
\end{equation}
Larger $\Delta_\ell$ indicates stronger parent--child coupling, and the protocol gaps are largest at such coupled depths, consistent with a decision-propagation bottleneck under JPD.

Overall, the protocol ladder attributes a substantial portion of the JPD drop to missing decision conditioning, especially at deeper, more coupled levels. Figure~\ref{fig:case} qualitatively illustrates the protocol gaps observed above: JPD can produce inconsistent paths, Independent Levels partially mitigates them, and DCR rollout further restores coherence via decision conditioning, consistent with our diagnosis.

\subsection{Mechanism Verification: Process vs.\ Content Supervision}
\label{sec:process_vs_content}

We test whether DuoTeach improves cross-level decision coordination by transferring a \emph{process} (decision-conditioned traces), rather than simply fitting taxonomy labels. We compare against a matched supervised fine-tuning baseline (SFT). Both methods train the same student under JPD with the same data and recipe; they differ only in supervision. SFT uses ground-truth level labels, while DuoTeach distills DCR-elicited teacher decisions and decision-point states.

On in-domain training taxonomies, both SFT and DuoTeach improve over Base under JPD, and their relative ranking depends on the backbone (Table~\ref{tab:sft_vs_sekd}). In contrast, on the unseen taxonomy Food-101, DuoTeach consistently outperforms SFT across all backbones, improving both path reliability and leaf accuracy (e.g., DWPA$_{0.95}$ gains of +4.06, +12.72, and +11.10 for LLaVA, Qwen, and InternVL3, respectively). This consistent advantage on an unseen taxonomy supports that distilling DCR decision traces transfers a more general cross-level coordination pattern, rather than relying on taxonomy-specific label fitting.

\begin{table}[t]
  \caption{SFT vs.\ DuoTeach on an in-domain taxonomy (Inat-Animal) and an unseen taxonomy (Food-101). All metrics are percentages (\%).}
  \label{tab:sft_vs_sekd}
  \centering
  \small
  \setlength{\tabcolsep}{6pt}
  \renewcommand{\arraystretch}{1.1}
  \begin{tabular}{l l cc cc}
    \toprule
    Model & Method & \multicolumn{2}{c}{In-domain} & \multicolumn{2}{c}{Food-101 (unseen)} \\
    \cmidrule(lr){3-4}\cmidrule(lr){5-6}
     &  & DWPA$_{0.95}$ & LeafAcc & DWPA$_{0.95}$ & LeafAcc \\
    \midrule
    \multirow{3}{*}{LLaVA-OV-7B}
      & Base     & 0.07 & 25.35 & 4.65 & 24.30 \\
      & SFT      & \textbf{6.56} & \textbf{41.80} & 40.37 & 82.20 \\
      & DuoTeach & 5.79 & 33.80 & \textbf{44.43} & \textbf{85.65} \\
    \midrule
    \multirow{3}{*}{Qwen2.5-VL-7B}
      & Base     & 10.41 & 44.15 & 17.17 & 59.85 \\
      & SFT      & \textbf{46.32} & \textbf{60.20} & 29.95 & 56.30 \\
      & DuoTeach & 21.23 & 48.75 & \textbf{42.67} & \textbf{90.25} \\
    \midrule
    \multirow{3}{*}{InternVL3-8B}
      & Base     & 3.11 & 30.15 & 8.33 & 27.00 \\
      & SFT      & 5.62 & 35.70 & 21.62 & 60.85 \\
      & DuoTeach & \textbf{17.73} & \textbf{41.75} & \textbf{32.72} & \textbf{68.40} \\
    \bottomrule
  \end{tabular}
\end{table}

\begin{table}[t]
  \caption{Ablation on distillation losses. All values are percentages (\%). Parentheses show absolute change vs.\ the full model. Full results are in Appendix G.}
  \label{tab:ablation_losses_transposed}
  \centering
  \begingroup
  \setlength{\tabcolsep}{4pt}
  \renewcommand{\arraystretch}{1.08}
  \small
  \begin{tabular}{@{}%
    l<{\hspace{4pt}} !{\color{black}\vrule width 0.3pt} >{\hspace{4pt}}c
    cc@{}}
    \toprule
    Variant & DWPA$_{0.95}$ & LeafAcc & TOR \\
    \midrule
    Base &
      3.11 &
      30.15 &
      27.64 \\
    Full DuoTeach &
      17.73 &
      41.75 &
      64.55 \\
    w/o $\mathcal{L}_{hard}$ &
      15.06 {\scriptsize(-2.67)} &
      41.30 {\scriptsize(-0.45)} &
      56.36 {\scriptsize(-8.19)} \\
    w/o $\mathcal{L}_{feat}$ &
      15.89 {\scriptsize(-1.84)} &
      42.65 {\scriptsize(+0.90)} &
      62.54 {\scriptsize(-2.01)} \\
    w/o $\mathcal{L}_{soft}$ &
      16.06 {\scriptsize(-1.66)} &
      43.85 {\scriptsize(+2.10)} &
      61.47 {\scriptsize(-3.08)} \\
    \bottomrule
  \end{tabular}
  \endgroup
\end{table}

\subsection{Ablation Study}
\label{sec:ablation}
We ablate the three loss terms used to distill DCR traces into a one-call student under JPD. We report DWPA$_{0.95}$ and TOR as primary path-level metrics, together with LeafAcc for reference.

Table~\ref{tab:ablation_losses_transposed} shows that removing any term degrades path reliability. Dropping $\mathcal{L}_{hard}$ causes the largest consistency drop, most notably in TOR (-8.19\,pp), indicating that matching the teacher's discrete decisions is crucial for maintaining coherent parent--child transitions. Removing $\mathcal{L}_{soft}$ or $\mathcal{L}_{feat}$ yields smaller but consistent reductions in DWPA$_{0.95}$ and TOR. LeafAcc may slightly increase when these regularizers are removed, suggesting that $\mathcal{L}_{soft}$ and $\mathcal{L}_{feat}$ mainly stabilize cross-level decision states rather than improving leaf recognition in isolation. Overall, DuoTeach performs best when combining label-level distillation with decision-point state alignment.

\newcommand{\smallparen}[1]{\textbf{\scriptsize\,(\,#1\,)}}
\newcommand{\valparen}[2]{%
  \begin{tabular}[c]{@{}c@{}}#1\\[-0.35ex]\smallparen{#2}\end{tabular}%
}

\begin{table*}[t]
  \caption{Generalization to an unseen taxonomy (Food-101) and external benchmarks. Food-101 is evaluated zero-shot under JPD (mean$\pm$std); parentheses on external benchmarks denote absolute gains (pp) over Base.
  All metrics are percentages (\%). Full results for additional base models are in Appendix F.2.}
  \label{tab:generalization}
  \centering
  \begingroup
  \setlength{\tabcolsep}{2pt}
  \renewcommand{\arraystretch}{1.0}
  \definecolor{RowBlue}{HTML}{E1E8FD}
  \definecolor{RowGray}{HTML}{E1E8FD}
  \small

  \begin{tabular}{@{}%
    >{\raggedright\arraybackslash}p{3.0cm}!{\color{black}\vrule width 0.3pt}
    c c!{\color{black}\vrule width 0.3pt}
    c!{\color{black}\vrule width 0.3pt}
    c c c@{}}
    \toprule

    \multicolumn{1}{l!{\color{black}\vrule width 0.3pt}}{\multirow{2}{*}[-0.55ex]{\textbf{Model}}} &
    \multicolumn{2}{c!{\color{black}\vrule width 0.3pt}}{\textbf{Food-101}} &
    \multicolumn{1}{c!{\color{black}\vrule width 0.3pt}}{\textbf{GQA}} &
    \multicolumn{3}{c}{\textbf{MathVista}} \\
    \cmidrule(lr){2-3}\cmidrule(lr){4-4}\cmidrule(lr){5-7}

    \multicolumn{1}{c!{\color{black}\vrule width 0.3pt}}{} &
    \multicolumn{1}{c}{\textbf{DWPA$_{0.95}$}} &
    \multicolumn{1}{c!{\color{black}\vrule width 0.3pt}}{\textbf{LeafAcc}} &
    \multicolumn{1}{c!{\color{black}\vrule width 0.3pt}}{\textbf{Acc}} &
    \multicolumn{1}{c}{\textbf{Avg.}} &
    \multicolumn{1}{c}{\textbf{Arith.}} &
    \multicolumn{1}{c}{\textbf{Geom.}} \\
    \midrule

    \multicolumn{1}{>{\columncolor{white}}l!{\color{black}\vrule width 0.3pt}}{InternVL3-8B \textit{Base}} &
    \multicolumn{1}{>{\columncolor{white}}c}{8.33} &
    \multicolumn{1}{>{\columncolor{white}}c!{\color{black}\vrule width 0.3pt}}{27.00} &
    \multicolumn{1}{>{\columncolor{white}}c!{\color{black}\vrule width 0.3pt}}{46.66} &
    \multicolumn{1}{>{\columncolor{white}}c}{64.90} &
    \multicolumn{1}{>{\columncolor{white}}c}{62.04} &
    \multicolumn{1}{>{\columncolor{white}}c}{70.29} \\

    \multicolumn{1}{>{\columncolor{RowBlue}}l!{\color{black}\vrule width 0.3pt}}{InternVL3-8B \textit{train}} &
    \multicolumn{1}{>{\columncolor{RowBlue}}c}{\valparen{23.88}{\,$\pm$9.46\,}} &
    \multicolumn{1}{>{\columncolor{RowBlue}}c!{\color{black}\vrule width 0.3pt}}{\valparen{53.62}{\,$\pm$14.56\,}} &
    \multicolumn{1}{>{\columncolor{RowBlue}}c!{\color{black}\vrule width 0.3pt}}{\valparen{53.05}{+6.39}} &
    \multicolumn{1}{>{\columncolor{RowBlue}}c}{\valparen{65.10}{+0.20}} &
    \multicolumn{1}{>{\columncolor{RowBlue}}c}{\valparen{62.61}{+0.57}} &
    \multicolumn{1}{>{\columncolor{RowBlue}}c}{\valparen{71.55}{+1.26}} \\

    \noalign{\vskip 0.5pt}\midrule\noalign{\vskip 0.5pt}

    \multicolumn{1}{>{\columncolor{white}}l!{\color{black}\vrule width 0.3pt}}{Qwen2.5-VL-7B \textit{Base}} &
    \multicolumn{1}{>{\columncolor{white}}c}{17.17} &
    \multicolumn{1}{>{\columncolor{white}}c!{\color{black}\vrule width 0.3pt}}{59.85} &
    \multicolumn{1}{>{\columncolor{white}}c!{\color{black}\vrule width 0.3pt}}{58.83} &
    \multicolumn{1}{>{\columncolor{white}}c}{64.80} &
    \multicolumn{1}{>{\columncolor{white}}c}{60.06} &
    \multicolumn{1}{>{\columncolor{white}}c}{63.18} \\

    \multicolumn{1}{>{\columncolor{RowGray}}l!{\color{black}\vrule width 0.3pt}}{Qwen2.5-VL-7B \textit{train}} &
    \multicolumn{1}{>{\columncolor{RowGray}}c}{\valparen{43.66}{\,$\pm$1.63\,}} &
    \multicolumn{1}{>{\columncolor{RowGray}}c!{\color{black}\vrule width 0.3pt}}{\valparen{90.09}{\,$\pm$0.35\,}} &
    \multicolumn{1}{>{\columncolor{RowGray}}c!{\color{black}\vrule width 0.3pt}}{\valparen{66.09}{+7.26}} &
    \multicolumn{1}{>{\columncolor{RowGray}}c}{\valparen{65.20}{+0.40}} &
    \multicolumn{1}{>{\columncolor{RowGray}}c}{\valparen{60.62}{+0.56}} &
    \multicolumn{1}{>{\columncolor{RowGray}}c}{\valparen{66.11}{+2.93}} \\
    \bottomrule
  \end{tabular}
  \endgroup
\end{table*}

\subsection{Generalization to Unseen Settings}
\label{sec:gemeralization_4_3}
We test whether DuoTeach transfers beyond the training taxonomies. We consider (i) zero-shot transfer to an unseen taxonomy (Food-101) and (ii) transfer to external benchmarks (GQA, MathVista).

\textit{Unseen taxonomy (Food-101).}
We train on five internal taxonomies and evaluate on Food-101 zero-shot under JPD. Table~\ref{tab:generalization} reports mean$\pm$std. DuoTeach substantially improves path reliability and leaf accuracy for both backbones (e.g., InternVL3-8B: DWPA$_{0.95}$ 8.33$\rightarrow$23.88; Qwen2.5-VL-7B: 17.17$\rightarrow$43.66), suggesting that the learned coordination is not tied to a single label tree.

\textit{External benchmarks.}
On GQA, DuoTeach improves Acc over the Base models (+6.39 and +7.26 pp), consistent with better within-call multi-step coordination. On MathVista, gains are smaller but consistent in overall accuracy (+0.20 and +0.40 pp), with clearer improvements on Arithmetic and Geometry.
Overall, these results are consistent with the interpretation that DuoTeach encourages an internalized stepwise policy:
the model becomes more likely to maintain intermediate decisions and reuse them when answering later sub-questions under single-invocation inference.

\begin{table}[t]
\caption{Model-size efficiency: a 7B DuoTeach student vs.\ a 32B Base model (Qwen2.5-VL) under JPD. All values are percentages (\%). Parentheses show 7B minus 32B (pp).}
  \label{tab:efficiency_7b_vs_32b}
  \centering
  \begingroup
  \setlength{\tabcolsep}{4pt}
  \renewcommand{\arraystretch}{1.08}
  \definecolor{HdrFill}{HTML}{E4E2FB}
  \small

  \begin{tabular}{%
    l<{\hspace{2pt}} !{\color{black}\vrule width 0.3pt}
    >{\hspace{2pt}}l<{\hspace{2pt}} !{\color{black}\vrule width 0.3pt}
    >{\hspace{2pt}}c<{\hspace{2pt}}
    >{\hspace{2pt}}c<{\hspace{2pt}}@{}}

    \toprule
    \multicolumn{1}{l!{\color{black}\vrule width 0.3pt}}{\cellcolor{HdrFill}\textbf{Dataset}} &
    \multicolumn{1}{>{\hspace{2pt}}l<{\hspace{2pt}}!{\color{black}\vrule width 0.3pt}}{\cellcolor{HdrFill}\textbf{Model}} &
    \multicolumn{1}{c}{\cellcolor{HdrFill}\textbf{DWPA$_{0.95}$}} &
    \multicolumn{1}{c}{\cellcolor{HdrFill}\textbf{LeafAcc}} \\
    \noalign{\vskip 1pt}\midrule\noalign{\vskip 1pt}

    \multirow{2}{*}[0pt]{iNat-Animal} &
      32B base & 22.87 & 55.55 \\
    & 7B student &
      21.23 \textbf{(-1.64)} &
      48.75 \textbf{(-6.80)} \\
    \noalign{\vskip 1pt}\midrule\noalign{\vskip 1pt}

    \multirow{2}{*}[0pt]{iNat-Plant} &
      32B base & 19.86 & 52.00 \\
    & 7B student &
      21.14 \textbf{(+1.27)} &
      48.95 \textbf{(-3.05)} \\
    \noalign{\vskip 1pt}\midrule\noalign{\vskip 1pt}

    \multirow{2}{*}[0pt]{ImgNet-Animal} &
      32B base & 49.93 & 78.20 \\
    & 7B student &
      50.31 \textbf{(+0.38)} &
      78.70 \textbf{(+0.50)} \\
    \noalign{\vskip 1pt}\midrule\noalign{\vskip 1pt}

    \multirow{2}{*}[0pt]{ImgNet-Artifact} &
      32B base & 18.62 & 83.00 \\
    & 7B student &
      13.17 \textbf{(-5.45)} &
      81.45 \textbf{(-1.55)} \\
    \noalign{\vskip 1pt}\bottomrule
  \end{tabular}
  \endgroup
\end{table}

\subsection{Practicality and Efficiency Analysis}
\label{sec:efficiency_analysis_4_5}
We assess practicality along two axes: model-size efficiency (a smaller student vs.\ a larger Base) and invocation efficiency (a one-call student vs.\ a multi-step DCR teacher).

\textit{Model-size efficiency.}
We compare a Qwen2.5-VL-7B DuoTeach student against the Qwen2.5-VL-32B Base model. As shown in Table~\ref{tab:efficiency_7b_vs_32b} and Table~\ref{tab:efficiency_summary}(a), the 7B student uses $4.0\times$ less peak VRAM (15.65\,GB) and achieves $2.81\times$ higher throughput at batch size $=1$, enabling deployment on 16--24\,GB GPUs. Despite its smaller scale, it remains competitive in path reliability on ImgNet-Animal and iNat-Plant, while the 32B Base is stronger on ImgNet-Artifact.

\textit{Invocation efficiency.}
DuoTeach eliminates the need for multi-step DCR at test time by distilling DCR traces into a one-call JPD student. Table~\ref{tab:efficiency_summary}(b) measures per-sample latency on iNat-Animal with InternVL3-8B, where DCR requires about $L\!\approx\!7$ invocations but JPD uses one. The student reduces latency by $3.34\times$ (1162.35 $\rightarrow$ 347.62 ms/sample).

Overall, DuoTeach preserves the JPD interface while improving path reliability and reducing both model footprint and invocation cost. We also observe no catastrophic forgetting on MMBench (Appendix F.2).

\begin{table}[t]
  \caption{Efficiency summary along two axes.
  (a) Model-size efficiency.
  (b) Invocation latency: multi-step DCR teacher vs.\ one-call JPD student (InternVL3-8B).}
  \label{tab:efficiency_summary}
  \centering
  \begingroup
  \setlength{\tabcolsep}{4pt}
  \renewcommand{\arraystretch}{1.06}
  \small

  \begin{tabular}{@{}l c c c@{}}

    \multicolumn{4}{c}{\textbf{(a) Model-size efficiency (batch size $=1$).}} \\
    \cmidrule(lr){1-4}
    Model & Params & Peak VRAM (GB) & Throughput (samples/sec) \\
    \midrule
    Base & 32B & 62.72 & 1.16 \\
    DuoTeach student & 7B & 15.65 & 3.26 \\

    \midrule
    \multicolumn{4}{c}{\textbf{(b) Invocation efficiency.}} \\
    \cmidrule(lr){1-4}
    Setting & Invocations & Latency (ms/sample) & Speedup \\
    \midrule
    DCR teacher & $L\approx 7$ & 1162.35 & $1.00\times$ \\
    JPD student & 1 & 347.62 & $3.34\times$ \\

    \bottomrule
  \end{tabular}

  \endgroup
\end{table}

\section{Conclusion}
\label{sec:conclusion}
We study coarse-to-fine path decision-making and introduce JPD with DWPA to evaluate one-call taxonomy-path reliability. Under JPD, strong VLMs frequently produce parent–child conflicts, revealing a cross-level decision-coordination bottleneck beyond missing taxonomic knowledge. DuoTeach uses a single pretrained VLM to elicit decision-conditioned traces via DCR and distill them into a one-call student, improving DWPA (including transfer to unseen taxonomies) while preserving deployment efficiency; future work extends this evaluation-and-self-teaching recipe to richer structures beyond explicit taxonomies and explores iterative refinement or integration with tool-augmented inference.

\bibliographystyle{splncs04}
\bibliography{main}
\end{document}